\documentclass{article}
\newcommand{\nn}{\nonumber\\}
\newcommand{\p}[1]{(\ref{#1})}

\newcommand{\cQ}{{\cal Q}}
\newcommand{\cQh}{\widehat{\cal Q}}

\newcommand{\ba}{\begin{array}}
\newcommand{\ea}{\end{array}}
\newcommand{\be}{\begin{equation}}
\newcommand{\ee}{\end{equation}}
\newcommand{\bea}{\begin{eqnarray}}
\newcommand{\eea}{\end{eqnarray}}
\newcommand{\bi}{\begin{itemize}}
\newcommand{\ei}{\end{itemize}}

\usepackage{amscd,amsmath,amssymb}
\begin{document}
\thispagestyle{empty}

\begin{center}
{~}\\
\vspace{3cm}
{\Large\bf N=8 Supersymmetric Quaternionic Mechanics}\\
\vspace{2cm}
{\large \bf S.~Bellucci${}^{a}$, S.~Krivonos${}^{b}$ and A.~Sutulin${}^{b}$ }\\
\vspace{2cm}
{\it ${}^a$INFN-Laboratori Nazionali di Frascati, C.P. 13,
00044 Frascati, Italy}\\
{\tt bellucci@lnf.infn.it} \\\vspace{0.5cm}
{\it ${}^b$ Bogoliubov  Laboratory of Theoretical Physics, JINR,
141980 Dubna, Russia}\\
{\tt krivonos, sutulin@thsun1.jinr.ru} \\ \vspace{2.5cm}
\end{center}

\begin{center}
{\bf Abstract}
\end{center}
We construct $N=8$ supersymmetric mechanics with four bosonic end eight fermionic
physical degrees of freedom. Starting from the most general $N=4$ superspace action in
harmonic superspace for the ({\bf 4,8,4}) supermultiplet we find conditions which make it
$N=8$ invariant. We introduce in the action Fayet-Iliopoulos terms which give rise
to potential terms. We present the action in components and give explicit expressions
for the Hamiltonian and Poisson brackets. Finally we discuss the possibility of $N=9$
supersymmetric mechanics.

\hfil
\newpage

\section{Introduction}
One dimensional extended supersymmetry has plenty of features which make it interesting not only
due to relations with higher dimensional theories but also as independent theory. The restrictions which
extended supersymmetry puts on the geometry of the target space (see e.g. \cite{VP1}) makes the systems with eight
real supercharges mostly interesting  among $N$ supersymmetric extended ones. Indeed, $N=8$ supersymmetry
is large enough to describe a rather complicated system and, at the same time, it does not put too strong
restrictions on the geometry of the bosonic sector.

In the last decade  some progress has been achieved in the construction of different variants of supersymmetric
mechanics with $N=8$ supersymmetry. In \cite{HKT} the structure of the sigma models with $N=4$ and $N=8$
supersymmetries has been analyzed. The first $N=8$ superconformally invariant action describing the low energy
effective dynamics of a D0-brane has been constructed in \cite{DE} (see also \cite{BMZ,Sm1,Sm2}).
Another version of $N=8$ superconformal mechanics with ({\bf 3,8,5}) and ({\bf 5,8,3}) $d=1$ supermultiplets
was constructed in \cite{BIKL1}. After a detailed investigation of the superfield structure of all possible $N=8, d=1$
supermultiplets in Ref. \cite{BIKL2},  new variants of $N=8$ supersymmetric mechanics with special K\"{a}hler
geometry in the bosonic sector were developed in \cite{BKN}.

The variety of $N=8, d=1$ supermultiplets contains a very special one -- i.e. the ({\bf 4,8,4}) supermultiplet. This
supermultiplet may be constructed by joining  two $N=4$ ones, namely ({\bf 4,4,0}) and ({\bf 0,4,4}) \cite{BIKL1}.
The supermultiplet ({\bf 4,4,0}) contains no auxiliary fields at all, while the ({\bf 0,4,4}) one does not
include any physical bosons. The detailed discussion of the $N=4$ ({\bf 4,4,0}) supermultiplet has been
performed in Ref. \cite{IL} within harmonic superspace. For $N=4$ supersymmetry this framework  is not the most general one,
because it puts additional constraint on the bosonic metric making the geometry strong HKT \cite{HKT}.
In the present paper we demonstrate that just this restriction is needed, in order to have $N=8$ supersymmetry.
We construct the most general superfields action for the ({\bf 4,8,4}) supermultiplet. We denote this system as
Supersymmetric Quaternionic Mechanics (SqM) because the physical bosons are represented just by quaternions $q^{ia}$.
We show that it is possible
to extend the action by a Fayet-Iliopoulos term, which results in the appearance of potential terms. We also provide
a complete Hamiltonian description of the constructed system. Finally, we proposed a $N=9$ supermultiplet and
prove that $N=9$ supersymmetry is too large to have an interesting action.

\section{N=8 SqM: Lagrangian}
In order to construct $N=8$ SqM with four bosonic and eight fermionic physical components we follow
\cite{BIKL2},
introducing a real quartet of $N=8$ superfields $\cQ^{ia}$  $\left( (\cQ^{ia})^\dagger = \cQ_{ia}\right)$ depending on
the coordinates of the
$N=8$, $d=1$ superspace $\mathbb{R}^{(1|8)}$
$$
\mathbb{R}^{(1|8)}=(t,\theta^{iA},\vartheta^{a\alpha})\,,\qquad
\left(\theta^{iA}\right)^\dagger=\theta_{iA}\,,\qquad
\left(\vartheta^{a\alpha}\right)^\dagger=\vartheta_{a\alpha}\,,
$$
where $i,\,a,\,A,\,\alpha=1,\,2$ are doublet indices of four commuting $SU(2)$ subgroups of
the automorphism group of $N=8$ Poincar\`{e} superalgebra\footnote{
We use the following convention for the skew-symmetric tensor $\epsilon$:
$\epsilon_{ij} \epsilon^{jk}=\delta_i^k$, $\epsilon_{12} = \epsilon^{21} =1.$}, and
obeying the constraints
\be\label{mc1}
D^{(i}_A \cQ^{j)a}=0, \quad \nabla^{(a}_\alpha \cQ^{b)i}=0\;.
\ee
The covariant spinor derivatives $D^{i}_A ,\nabla^{a}_\alpha$ are defined by
\bea\label{sderiv}
&& D^{iA}=\frac{\partial}{\partial\theta_{iA}}+i\theta^{iA}\partial_t\,,\quad
\nabla^{a\alpha}=\frac{\partial}{\partial\vartheta_{a\alpha}}+
i\theta^{a\alpha}\partial_t\,, \\
&& \left\{ D^{iA},D^{jB}\right\}=2i \epsilon^{ij}\epsilon^{AB}\partial_t, \quad
\left\{ \nabla^{a\alpha},\nabla^{b\beta}\right\}=2i \epsilon^{ab}\epsilon^{\alpha\beta}\partial_t,\quad
\left\{ D^{iA},\nabla^{a\alpha}\right\}=0\;. \nonumber
\eea
Using \p{mc1}, it is possible to show that the superfields $\cQ^{ia}$ contain the following
independent components:
\be\label{comp}
q^{ia}=\cQ^{ia}|,\quad \psi^{aA}=\frac{1}{2}D^A_i\cQ^{ia}|,\quad \xi^{i\alpha}=\frac{1}{2}\nabla^\alpha_a\cQ^{ia}|,\quad
F^{\alpha A}=D^A_i \nabla^\alpha_a \cQ^{ia}|\;,
\ee
where $|$ denotes the restriction to $\theta_{iA}=\vartheta_{a\alpha}=0$. Thus, we deal with the irreducible
({\bf 4,8,4}) supermultiplet.

In order to construct the corresponding action, we pass to $N=4$ superfields.
As a first step, we single out
a $N=4$ subspace in the $N=8$ superspace $\mathbb{R}^{(1|8)}$ as the set of coordinates
\be
\mathbb{R}^{(1|4)} = \left( t, \theta_{iA} \right) \subset \mathbb{R}^{(1|8)} \label{ss1}
\ee
and expand the $N=8$ superfields over the additional Grassmann coordinate $\vartheta_{a\alpha}$.
Due to the following corollary of the constraints \p{mc1}:
\be\label{corol}
\nabla^{a\alpha}\nabla^{b\beta} \cQ^{ic}=2i\epsilon^{\alpha\beta}\epsilon^{cb}{\dot{\cQ}}{}^{ia},
\ee
one may easily see that in this expansion only the following four bosonic and four fermionic $N=4$
superfield projections:
\be
q^{ia}=\cQ^{ia}|_{\vartheta=0},\quad \xi^{i\alpha}=\frac{1}{2}\nabla^\alpha_a \cQ^{ia}|_{\vartheta=0}
\ee
are independent.
These $N=4$ superfields are subjected, in virtue of
eqs. \p{mc1}, to the irreducibility constraints in $N=4$
superspace
\be\label{mc2}
D_A^{(i}q^{j) a}=0,\quad
D_A^{(i} \xi^{j) \alpha}=0\,.
\ee
The implicit $N=4$ Poincar\'e supersymmetry transformations,
completing the manifest one to the full $N=8$ supersymmetry, have the
following form in terms of these $N=4$ superfields:
\be\label{tr1}
\delta q^{ia}=-\varepsilon^{a\alpha} \xi^{i}_\alpha,\quad
\delta \xi^{i\alpha}=2i\varepsilon^{a\alpha} {\dot q}{}_a^i.
\ee

The simplest way to deal with an action for the supermultiplet $\cQ^{ia}$
is to use the harmonic superspace approach \cite{{harm},{book},{IL}}.
We use the definitions and conventions of Ref. \cite{IL}. The harmonic
variables parameterizing  the coset $SU(2)_R/U(1)_R$
are defined by the relations
\be\label{h1}
u^{+i}u^-_i = 1 \quad \Leftrightarrow\quad
u^+_i u^-_j - u^+_j u^-_i = \epsilon_{ij} \;, \;\;
\overline{(u^{+i})} = u^-_i\,.
\ee
The harmonic projections of
$q^{ia}$ and $\xi^{i\alpha}$ are defined by
\be
q^{+a}=q^{ia} u^+_i,\quad  \xi^{+\alpha}=\xi^{i\alpha} u^+_i,\;\label{h2}
\ee
and the constraints \p{mc2} are rewritten as
\be\label{mc3}
D^{+A}q^{+a}=D^{+A}\xi^{+\alpha}=0,\quad D^{++}q^{+a}=D^{++}\xi^{+\alpha}= 0\;.
\ee
Here $D^{\pm A}= D^{iA}u_i^\pm$ and
$D^{\pm\pm} = u^{\pm i}\partial/\partial u^{\mp i}$ (in the central
basis of the harmonic superspace), with $D^{iA} $ given in \p{sderiv}.

The relations \p{mc3} imply that
$q^{+a},\xi^{+\alpha}$ are constrained analytic harmonic $N=4, d=1$ superfields
living on the analytic subspace $(\zeta, u^\pm_i)
\equiv (t_A, \theta^{+A}, u^\pm_i)$ that is closed under
$N=4$ supersymmetry. In this setting, the  hidden
$N=4$ supersymmetry  transformations \p{tr1} are rewritten as
\be\label{tr2}
\delta q^{+a}=-\varepsilon^{a\alpha} \xi^{+}_\alpha,\quad
\delta \xi^{+\alpha}=2i\varepsilon^{a\alpha} {\dot q}{}_a^+.
\ee

The generic action has the form
\be\label{action1}
S= \int du d\zeta^{--}\Big \{ {\cal L}^{+ a}(q^{+ b},u)\, \partial_t\, q^+_a
+ {\cal L}(q^{+ b},u)\, \xi^{+ \alpha}\, \xi^+_\alpha
\,\Big\}
\ee
where ${\cal L}^{+ a}(q^{+ b},u)$ and ${\cal L}(q^{+ b},u)$ are, for the time being, arbitrary
functions of $q^{+ a}$ and the harmonics $u$, and
$du d\zeta^{--}= du dt_A d\theta^{+B} d\theta^+_B $
is the measure of integration over the analytic superspace.
The action \p{action1} is manifestly $N=4$ supersymmetric since
it is written in terms of $N=4$ superfields.
However, its invariance with respect to the hidden $N=4$ supersymmetry
\p{tr2} must be explicitly checked. Varying the integrand in \p{action1} with respect to
\p{tr2} one may find that, in order to have  hidden $N=4$ supersymmetry, one should impose
the following condition:
\be \label{n8inv}
\frac{\partial {\cal L}^{+ a}}{\partial q^{+a}} = 4i\,{\cal L}\,.
\ee
The immediate corollary is that {\it any action} written in terms of the $N=4$
superfields $q^{ia}$ in the {\it harmonic superspace} can be promoted to an invariant of $N=8$
supersymmetry by adding the interaction with the superfields $\xi^{i \alpha}$.
We would like to stress that the most general $N=4$ supersymmetric action for the superfields
$q^{ia}$ subjected to the constraints \p{mc2} should be written in the full $N=4$ superspace as
\be\label{genaction}
S=\int dt d^4\theta L(q) \;,
\ee
where $L(q)$ is an arbitrary scalar function. Requiring the $N=4$ supersymmetric action for $q^{ia}$
to be representable in the analytic harmonic superspace imposes severe restrictions on the
target-space geometry \cite{IL}. To clarify these restrictions and their consequences for the
$N=8$ case let us pass to the components action. Integrating in \p{action1} (with the constraints \p{n8inv} imposed)
over Grassmann variables and
eliminating the auxiliary fields $F^{a\alpha}$ by their equations of motion, we end up with the
following action:
\bea\label{action2}
S&=& \int dt\, \left[ G \left( \dot{q}^{ia}\dot{q}_{ia}
+ \frac{i}{2}\xi^{i\alpha}\dot{\xi}_{i\alpha}
+ \frac{i}{2}\psi^{aA}\dot{\psi}_{aA}\right)+
\frac{i}{2} \frac{\partial G}{\partial q^{ia}} \left(  \xi^{i\alpha}\xi_\alpha^k\; \dot{q}^a_k
+ \psi^{aA}\psi_A^b\; \dot{q}^i_b\right) \right. + \nn
&& \left. \frac{1}{8} \left( \frac{\partial^2 G}{\partial q^{ia} \partial q^{kb}}
- 2 G^{-1}\,\frac{\partial G}{\partial q^{ia}} \frac{\partial G}{\partial q^{kb}}\right)\,
 \xi^{i\alpha}\xi_\alpha^k\;\psi^{aA}\psi_A^b\right] \;,
\eea
where the metric is defined as
\be\label{metric}
G = \int du\, \left.{\cal L}(q^{+ i},u) \right|_{\theta = 0} \;.
\ee
It follows immediately  from \p{metric} that such a metric $G$ satisfies the condition \cite{IL}
\be\label{hkt}
\bigtriangleup G\equiv \frac{\partial^2}{\partial q^{ia} \partial q_{ia}}\;G=0 \;.
\ee
This type of metrics defines a strong HKT geometry \cite{HKT}.

It is worth pointing out that the restriction \p{metric} in the $N=4$ case makes the theory almost
trivial. Indeed, passing to the components in the most general action \p{genaction} we  get
\be
S_{N=4}=\int dt \left[ G \left( \dot{q}^{ia}\dot{q}_{ia}
+ \frac{i}{2}\psi^{aA}\dot{\psi}_{aA}\right)+
\frac{i}{2} \frac{\partial G}{\partial q^{ia}} \psi^{aA}\psi_A^b\; \dot{q}^i_b -
\frac{1}{3}\bigtriangleup G\,\psi^{aA}\psi_A^b \psi^{B}_a\psi_{Bb} \right],
\ee
where
\be
G=\frac{\partial^2}{\partial q^{ia} \partial q_{ia}}L(q) \;.
\ee
Now, it is evident that the constraint \p{hkt}, being imposed, kills the four-fermion
interaction term in the action. Correspondingly, the Hamiltonian for such a theory does not contain
fermions
at all and  describes the  pure bosonic sigma-model. In the $N=8$ SqM case
the action \p{action2} still contains a four-fermion term which
combines the fermions coming from two $N=4$ supermultiplets $q^{ia}$ and $\xi^{i\alpha}$.

Thus, we conclude that the action \p{action2} is invariant with respect to $N=8$ supersymmetry, which is realized
on the physical component fields as follows:
\be\label{tr2c}
\delta q^{ia}=-\epsilon^{iA}\psi^a_A -\varepsilon^{a\alpha}\xi^i_\alpha\;,\quad
 \delta\psi^{aA}=2i\epsilon^{iA}{\dot q}_i^a -\varepsilon^{a\alpha}F_\alpha^A\;, \quad
\delta\xi^{i\alpha}=\epsilon^{iA}F^\alpha_A+2i\varepsilon^{a\alpha}{\dot q}^i_a\;,
\ee
with $\epsilon^{iA}$, $\varepsilon^{a\alpha}$ being the parameters of two $N=4$
supersymmetries acting on
$\theta^{iA}$ and $\vartheta^{a\alpha}$, respectively, and with the auxiliary
fields $F_{\alpha A}$ defined as
\be
F_{\alpha A} = -G^{-1}\frac{\partial G}{\partial q^{ia}}\xi^i_\alpha \psi^a_A \;.
\ee
Using the Noether theorem one can find classical expressions for the conserved
supercharges
corresponding to the supersymmetry transformations \p{tr2c}
\be\label{Q}
Q^{iA}=G\, \dot{q}{}^i_a\, \psi^{aA}+
\frac{i}{3}\frac{\partial G}{\partial q^b_i}\,\psi^{Ac}\,\psi_{cD}\,\psi^{bD}\;,   \quad
S^{a\alpha}=G\, \dot{q}{}^a_i\, \xi^{i\alpha}+
\frac{i}{3}\frac{\partial G}{\partial q_a^k}\,\xi^{\alpha}_j\,\xi^{j\beta}\,\xi^{k}_\beta\;.
\ee

We conclude this section with a few comments.

Firstly, we succeeded in the construction of a $N=8$ supersymmetric action for the ({\bf 4,8,4})
supermultiplet \p{action2} with the
bosonic metric $G$ subjected to the constraint \p{hkt}. This restriction appears naturally in such
consideration because the model
formulates in the harmonic superspace. In the next Section we will demonstrate
that this constraint is not
an {\it artifact} of this approach, but it is an unavoidable feature of $N=8$ SqM.

Secondly, the action \p{action1} may be extended by the Fayet-Iliopoulos term
\be
\widetilde{S}=S+  \int du d\zeta^{--}\,\lambda_{A\alpha}\, \theta^{+A}\,\xi^{+\alpha}\;,
\ee
with  constant real matrix parameters $\lambda^{A\alpha}$
($\left( \lambda^{A\alpha}\right)^\dagger =\lambda_{A\alpha}$).
As a result, the component action gets the potential terms
\be
S_p=\frac{1}{16}\int dt\left[ \frac{\lambda^2}{G}+4\, G^{-1}\,
\frac{ \partial G}{\partial q^{ia}}\,\lambda _{A\alpha}\,\xi^{i\alpha}\,\psi^{aA}\right].
\ee
Such a possibility to get potential terms in the joint action for ({\bf 4,4,0}) and ({\bf 0,4,4})
supermultiplets has been noted in Ref. \cite{IL}.

A final comment concerns the $N=8$ superconformal invariant action. From the considerations in
Refs. \cite{BIKL1,BIKL2} it follows that
the corresponding $N=8$ superconformal group should be the $OSp(4^\star|4)$ one.
As an additional requirement, the bosonic metric
$G$ should be invariant with respect to all automorphism transformations,
which means $G=G(q^2)$. Thus, as a solution of the
constraint \p{hkt}, we have
\be
G_{conf}=\frac{a}{q^2} +b,\quad  a,b =\mbox{const}.
\ee
The solution with $a\neq 0$ explicitly breaks even dilatation invariance.
So, we conclude that the unique candidate to be
a $N=8$ superconformal action is a free action with constant metric $G$.

\section{N=8 SqM: Hamiltonian}
In order to find the classical Hamiltonian, we  define the momenta $p_{ia}$, $\pi^{(\psi)}_{aA}$,
$\pi^{(\xi)}_{a\alpha}$ conjugated to
$q^{ia}$, $\psi^{aA}$ and $\xi^{i\alpha}$, respectively, as
\be\label{momenta}
p_{ia}=2\,G\,{\dot q}_{ia}-\frac{i}{2}\frac{\partial G}{\partial q^{ib}}\psi^{aA}\psi^b_A -
 \frac{i}{2}\frac{\partial G}{\partial q^{ja}}\xi^{i\alpha}\xi^j_\alpha\;, \quad
\pi^{(\psi)}_{aA} =-\frac{i}{2}\,G\,\psi_{aA},\; \pi^{(\xi)}_{i\alpha} =-\frac{i}{2}\,G\,\xi_{i\alpha}
\ee
and introduce the canonical Poisson brackets
\be\label{pb1}
\left\{ q^{ia},p_{jb} \right\}=\delta^i_j \delta^a_b,\quad
\left\{ \psi^{aA} ,\pi^{(\psi)}_{bB} \right\}=-\delta^a_b \delta^A_B,\quad
\left\{ \xi^{i\alpha} ,\pi^{(\psi)}_{j\beta} \right\}=-\delta^\alpha_\beta \delta^i_j.
\ee
{}From the explicit form of the fermionic momenta \p{momenta} it
follows that the system possesses  second--class constraints. Using the standard procedure,
we get the following Dirac brackets for the canonical variables\footnote{From now on, the symbol
$\{\,,\}$ stands for the Dirac bracket.} :
\bea\label{pb2}
&&
\left\{ q^{ia},p_{jb} \right\}=\delta^i_j \delta^a_b,\quad
\left\{ \psi^{aA} ,\pi^{(\psi)}_{bB} \right\}=-\delta^a_b \delta^A_B,\quad
\left\{ \xi^{i\alpha} ,\pi^{(\psi)}_{j\beta} \right\}=-\delta^\alpha_\beta \delta^i_j \;,\nn
&&
\{\tilde p^{ia}, \tilde p^{jb}\} = - \frac{i}{2}\,
\epsilon^{ab}R^i{}_c{}^j{}_d \,\psi^{cA}\psi_A^d  - \frac{i}{2}\,
\epsilon^{ij}R_m{}^a{}_n{}^b\, \xi^{m\alpha}\xi_\alpha^n  \nn
&&
\{\tilde p^{ia}, \psi_{bA}\} = \delta^a_b\,\Gamma^{id}\,\psi_{dA}\,, \quad
\{\tilde p^{ia}, \xi_{k\beta}\} = \delta^i_k\,\Gamma^{la}\,\xi_{l \beta}\,, \nn
&&
\{\psi^{aA}, \psi^{bB}\} = - \frac{i}{G}\, \epsilon^{ab}\,\epsilon^{AB}\,, \quad
\{\xi^{i \alpha}, \xi^{k \beta}\} = - \frac{i}{G}\, \epsilon^{ik}\,\epsilon^{\alpha \beta}\,,
\eea
where
\be\label{defR}
R_{iajb} =\frac{\partial^2 G}{\partial q^{ia} \partial q^{jb}}-
\frac{2}{G}\frac{\partial G}{\partial q^{ia}}\frac{\partial G}{\partial q^{jb}},\quad
\Gamma_{i\,a} =\frac{1}{2}\, \frac{\partial \ln G}{\partial q^{i\,a}}
\ee
and the bosonic momenta ${\tilde p}{}^{ia}$ have been defined as
\be
\tilde p^{i\,a} = p^{i\,a} +  i\, G\,\left( \Gamma^a_k\,  \xi^{i\alpha}\xi_\alpha^k
+ \Gamma^i_b\, \psi^{aA}\psi_A^b
\right).
\ee
Now one can check that the supercharges $Q_{iA}$, $S_{a\alpha}$ \p{Q}, being rewritten through
the momenta ${\tilde p}{}^{ia}$ as
\be\label{Q1}
Q^{iA}={\tilde p}{}^i_a\, \psi^{aA}+\frac{i}{3}\frac{\partial G}{\partial q^b_i}\,\psi^{Ac}\,\psi_{cD}\,\psi^{bD}\;,   \quad
S^{a\alpha}={\tilde p}{}^a_i\, \xi^{i\alpha}+\frac{i}{3}\frac{\partial G}{\partial q_a^k}\,\xi^{\alpha}_j\,\xi^{j\beta}\,\xi^{k}_\beta\;,
\ee
and the Hamiltonian
\be\label{ham}
H=\frac{{\tilde p}{}^2}{2G} - \frac{1}{4}R_{iajb}\,\psi^{aA}\psi^b_A\, \xi^{i\alpha}\xi^j_\alpha
\ee
form the standard $N=8$ superalgebra:
\be\label{SA}
\{ Q^{iA},  Q^{kB}\} = -i\,\epsilon^{ik}\,\epsilon^{AB}\, H\,,\quad
\{ S^{a\alpha}, S^{b\beta}\} =
- i\,\epsilon^{ab}\,\epsilon^{\alpha\beta}\,  H\,,\quad
\{  Q^{iA},   S^{a\alpha}\} = 0\,.
\ee
With these, we completed the classical description of $N=8$ SqM. Before closing this Section, let us
go back and clarify the  necessities of the constraint \p{hkt}. The idea is  rather simple. The Dirac brackets \p{pb2}
provide a quite general basis to construct the system without any constraints on the metric. The most general
Ansatz for the supercharges $Q^{iA}$ and $S^{a\alpha}$ reads
\be\label{newQ}
Q^{iA} = {\tilde p}^i_a\, \psi^{aA} + f^i_a\, \psi^{aB}\psi_{dB}\psi^{dA}
+ g_{ka}\,\psi^{aA}\, \xi^{k \alpha}\xi^i_{\alpha}\,,\;
S^{a\alpha} ={\tilde p}^a_i\,\xi^{i\alpha}
+ t^a_i\,\xi^i_{\beta}\xi^{k\alpha}\xi_k^{\beta}
+ h_{ib}\,\xi^{i\alpha}\psi^{aA}\psi^b_A\;,
\ee
where $f^i_a, g_{ka}, t^a_i$ and $h_{ib}$ are arbitrary functions of $q^{ia}$.
Straightforward but rather lengthly calculations show that the supercharges \p{newQ}
obey the superalgebra \p{SA} only if they have form as in \p{Q1} and if the metric $G$ obeys
\p{hkt}. So, the main restriction on the metric \p{hkt} is an unavoidable feature of $N=8$ SqM.

Finally, let us note that the generalization of  $N=8$ SqM to the $n$-dimensional case is straightforward
and will be presented elsewhere.

\section{Towards N=9 supersymmetry}
One of the interesting features of the $N=8$ supermultiplets with ({\bf 4,8,4}) content is the possibility
to realize one additional supersymmetry on a pair of such multiplets. This can be done as follows. Let us
start with a pair of $N=8$ supermultiplets $\cQ{}^{ia}$ and $\cQh{}^{i\alpha}$ subject to the constraints
\be\label{mc9}
D^{(i}_A \cQ^{j)a}=0, \quad \nabla^{(a}_\alpha \cQ^{b)i}=0\;, \qquad
D^{(i}_A \cQh^{j)\alpha}=0, \quad \nabla^{(\alpha}_a \cQh^{\beta)i}=0\;.
\ee
In the $N=4$ subspace of  the $N=8$ superspace $\mathbb{R}^{(1|4)}$ \p{ss1}
only the following eight bosonic and eight fermionic $N=4$
superfields projections:
\be
q^{ia}=\cQ^{ia}|_{\vartheta=0},\quad \xi^{i\alpha}=\frac{1}{2}\nabla^\alpha_a \cQ^{ia}|_{\vartheta=0},\qquad
{\hat q}^{i\alpha}=\cQh{}^{i\alpha}|_{\vartheta=0},\quad \hat\xi{}^{ia}=\frac{1}{2}\nabla^a_\alpha \cQh^{i\alpha}|_{\vartheta=0},
\ee
are independent.
These $N=4$ superfields are subject, in virtue of
eqs. \p{mc9}, to the irreducibility constraints in $N=4$
superspace
\be\label{mc92}
D_A^{(i}q^{j) a}=0,\;
D_A^{(i} \xi^{j) \alpha}=0\,,\qquad D_A^{(i}{\hat q}{}^{j) \alpha}=0,\;
D_A^{(i} \hat\xi{}^{j) a}=0\,.
\ee
The transformations under implicit $N=4$ supersymmetry have the form
\be\label{tr92}
\delta q^{ia}=-\eta^{a\alpha} \xi^{i}_\alpha,\;
\delta \xi^{i\alpha}=2i\eta^{a\alpha} {\dot q}{}_a^i,\quad
\delta {\hat q}{}^{i\alpha}=-\eta^{a\alpha} \hat\xi^{i}_a,\;
\delta \hat\xi{}^{ia}=2i\eta^{a\alpha} {\dot q}{}_\alpha^i,
\ee
while the additional, ninth supersymmetry may be realized on these $N=4$ superfields as
\be\label{n9}
\delta q^{ia} = - \epsilon\, \hat\xi^{ia}\,, \quad
\delta \xi^{a\alpha} = -2i\, \epsilon\, \partial_t {\hat q}^{a\alpha}\,, \qquad
\delta {\hat q}{}^{a\alpha} = - \epsilon\, \xi^{a\alpha}\,, \quad
\delta \hat\xi{}^{ia} = -2i\, \epsilon\, \partial_t\, q^{ia}\,.
\ee
Thus, we have a $N=9$ supermultiplet with ({\bf 8,16,8}) components in full agreement with
the considerations in Refs. \cite{GR}, where it has been proven that the minimal $N=9$ supersymmetric
multiplet must have 16 bosonic and 16 fermionic fields.

As in the case of one $N=8$ supermultiplet $\cQ{}^{ia}$ we considered in the previous Sections,
the harmonic superspace provides the most adequate setup for constructing the action.
After introducing harmonic projections of
$q^{ia},\xi^{i\alpha},{\hat q}{}^{i\alpha}$ and $\hat\xi{}^{ia}$ by
\be
q^{+a}=q^{ia} u^+_i,\;  \xi^{+\alpha}=\xi^{i\alpha} u^+_i,\qquad
{\hat q}{}^{+\alpha}={\hat q}{}^{i\alpha} u^+_i,\;  \hat\xi{}^{+a}=\hat\xi{}^{ia} u^+_i,
\label{h92}
\ee
one may write the most general Ansatz for the hypothetical $N=9$ supersymmetric action
\bea\label{action9}
S_{N=9} &=&  \int du d\zeta^{--}\, \Big \{
{\cal L}^{+ a}(q^{+ b},{\hat q}{}^{+\beta},u)\, \partial_t\, q^+_a
+ {\cal L}^{+ \alpha}(q^{+ b},{\hat q}{}^{+\beta},u)\, \partial_t\, {\hat q}{}^+_\alpha \nn
 &+&  {\cal L}(q^{+ b},{\hat q}{}^{+\beta},u)\, \xi^{+ \alpha}\, \xi^+_\alpha
+ \tilde {\cal L}(q^{+ b},{\hat q}{}^{+\beta},u)\, \hat\xi{}^{+ a}\, \hat\xi{}^+_a
+ {\cal L}_{\alpha a}(q^{+ b},{\hat q}{}^{+\beta},u)\, \xi^{+ \alpha}\, \hat\xi{}^{+ a}
\,\Big\},
\eea
where ${\cal L}^{+ a},{\cal L}^{+ \alpha},{\cal L},\tilde {\cal L},{\cal L}_{\alpha a}$ are arbitrary
functions of $N=4$ superfields $q^{+a}, {\hat q}{}^{+\alpha}$ and harmonics $u^\pm_i$.

Being written in $N=4$ superspace, the action \p{action9} enjoys manifest $N=4$ supersymmetry while the
invariance with respect to implicit $N=4$ \p{tr92} and ninth supersymmetry \p{n9} should be checked.
The results of  rather length calculations are
\begin{itemize}
\item{\bf N=8}: the invariance with respect to implicit $N=4$ supersymmetry \p{tr92} and, therefore the
$N=8$ invariance of the action put the following restrictions:
\bea
&& \frac{\partial {\cal L}^{+a}}{\partial {\hat q}{}^{+\alpha}}=\frac{\partial {\cal L}}{\partial {\hat q}{}^{+\alpha}}=0,\quad
\frac{\partial {\cal L}^{+\alpha}}{\partial q^{+a}}=\frac{\partial \tilde{\cal L}}{\partial q^{+a}}=0,\quad
{\cal L}_{\alpha a}=0, \label{a} \\
&&\frac{\partial {\cal L}^{+ a}}{\partial q^{+ a}} = 4i\, {\cal L}\,, \quad
\frac{\partial {\cal L}^{+ \alpha}}{\partial {\hat q}^{+\alpha}} = 4i\, \tilde {\cal L}\,. \label{b}
\eea
Thus, $N=8$ supersymmetry forbids the interaction between two supermultiplets $q^{ia},\xi^{i\alpha}$ and
${\hat q}{}^{i\alpha},\hat\xi{}^{ia}$. This result is not so strange and may be treated as a dimensionally reduced
version of the results obtained in Ref. \cite{ISut}, where possible interactions of hypermultiplets in $d=2$
have been analyzed.
\item{\bf N=9}: the ninths supersymmetry makes the action completely free
\be
{\cal L}^{+a}=q^{+a},\quad {\cal L}^{+\alpha}={\hat q}^{+\alpha}.
\ee
\end{itemize}
Therefore, we conclude that $N=9$ supersymmetry is so restrictive that only free actions can enjoy it.
These results may be considered as an indirect proof (not complete, of course) of the statement that
the theories with eight supercharges are the
highest $N$  theories with extended supersymmetries which have a rich geometric structure
of the target space (see e.g. \cite{VP1}).

\section{Conclusions} In this paper we constructed a new version of $N=8$ supersymmetric mechanics
based on the off-shell multiplet ({\bf 4,8,4}). We showed that the most general action for this supermultiplet
can be constructed within harmonic superspace. We extended the action by a Fayet-Iliopoulos term, which gives
rise to potential terms.  We also provide
a detailed Hamiltonian description of the constructed system.

The supermultiplet ({\bf 4,8,4}) is  rather unique, because it may be decomposed into the two very special
$N=4$ supermultiplets ({\bf 4,4,0}) and ({\bf 0,4,4}), which cannot exist in higher dimensions. As an
interesting result, we get the restriction on the bosonic metric, which should obey the Laplace equation.
This restriction, which is automatically satisfied in the harmonic superspace approach for the ({\bf 4,4,0})
supermultiplet \cite{IL} and appears to be too strong in $N=4$ supersymmetric mechanics, is unavoidable
in the case of $N=8$ supersymmetry. Combining two ({\bf 4,8,4}) supermultiplets we constructed a $N=9$ supermultiplet.
Unfortunately, $N=9$ supersymmetry is too large and restricts the most general action to be a free one.

Apart form the considered decomposition of the $N=8$ supermultiplet ({\bf 4,8,4}) into ({\bf 4,4,0}) and ({\bf 0,4,4}),
there is another one ({\bf 4,8,4})=({\bf 2,4,2})+({\bf 2,4,2}) \cite{BIKL2}. It seems interesting to find
appropriate Lagrangian and Hamiltonian descriptions of $N=8$ SqM for such a splitting, especially due to the existence
of a complete Hamiltonian analysis of the $N=4$ supersymmetric mechanics with 4n+4n phase space \cite{BN}.

An obvious project for future study is to formulate $N=8$ SqM in the $N=8$ harmonic superspace with double
sets of harmonics, similarly to Ref. \cite{ISut2}. Another possible extension of the proposed system is related
with the relaxing of the constraints for the ({\bf 0,4,4}) supermultiplet by admitting a constant part for the
auxiliary fields. As a result, in such a version we  expect the appearance of  central charges in the
$N=8$ super Poincar\'{e} algebra.

\section*{Acknowledgements}
We wish to thank Armen Nersessian for many enlightening comments and discussions.

This research was partially supported by the European
Community's Marie Curie Research Training
Network under contract MRTN-CT-2004-005104 Forces Universe,
INTAS-00-00254 grant,
RFBR-DFG grant No 02-02-04002, grant DFG No 436 RUS 113/669, and RFBR grant
No 03-02-17440.
S.K. thanks INFN --- Laboratori Nazionali di Frascati  for the warm
hospitality extended to him during the course of this work.

\end{document}